\documentclass[11pt]{article}
\usepackage{cite}
\usepackage{amsmath,amsfonts,amssymb}
\usepackage[small,bf,hang]{caption}
\usepackage{slashed}

\def\hybrid{
        \topmargin -20pt
        \oddsidemargin 0pt
        \headheight 0pt \headsep 0pt
        \textwidth 6.25in 
        \textheight 9.5in 
        \marginparwidth .875in
        \parskip 5pt plus 1pt \jot = 1.5ex}

\hybrid

 \csname
@addtoreset\endcsname{equation}{section}

\newcommand{\vev}[1]{\langle #1 \rangle}

\def\moth{\mathsurround=0pt}
\newdimen\zo \zo=0pt

\def\tick{\leaders\hrule height 0.5ex depth 0pt \hskip 0.5pt}
\def\upboxfill{$\moth \setbox\zo\hbox{\tick}%
  \hskip 3pt\hbox to 0pt{$\tick$\hss}\hrulefill \hbox to 7.5pt{$\tick$\hss}$}

\def\dtick{\leaders\hrule height .34pt depth 0.5ex \hskip 0.5pt}
\def\downboxfill{$\moth \setbox\zo\hbox{\dtick}%
  \hskip 2pt\hbox to 0pt{$\dtick$\hss}\hrulefill \hbox to 2pt{$\dtick$\hss}$}

\def\bec{\begin{center}}
\def\ec{\end{center}}

 \def\det{{\rm det\,}}
\def\be{\begin{equation}}
\def\ee{\end{equation}}
\def\bea{\begin{eqnarray}}
\def\eea{\end{eqnarray}}
\def\ba{\begin{array}}
\def\ea{\end{array}}

\def\ket#1{|#1\rangle}

\def\fancys{\mathbb{S}}

\begin{document}

\begin{titlepage}
\rightline{}
\rightline{\tt MIT-CTP-4288}
\rightline{August 2011}
\begin{center}
\vskip 2.5cm
{\Large \bf {
Massive Type II in Double Field Theory
}}\\
\vskip 3.0cm
{\large {Olaf Hohm and Seung Ki Kwak}}
\vskip 0.5cm
{\it {Center for Theoretical Physics}}\\
{\it {Massachusetts Institute of Technology}}\\
{\it {Cambridge, MA 02139, USA}}\\
ohohm@mit.edu, sk\_kwak@mit.edu
\vskip 0.7cm

\vskip 1.5cm
{\bf Abstract}
\end{center}

\vskip 0.5cm

\noindent
\begin{narrower}
We provide an extension of the recently constructed double field theory
formulation of the low-energy limits of type II strings, in which the RR fields
can depend simultaneously 
on the 10-dimensional space-time coordinates
and linearly on the dual winding coordinates. For the special case that only the
RR one-form of type IIA carries such a dependence,
we obtain the massive deformation of type IIA supergravity due to Romans.
For T-dual configurations we obtain a `massive' but non-covariant
formulation of type IIB, in which the 10-dimensional diffeomorphism symmetry
is deformed by the mass parameter.

\end{narrower}

\end{titlepage}

\newpage

\section{Introduction}
Double field theory is an approach to make T-duality manifest at the level
of effective space-time theories by doubling the coordinates.  It introduces
the usual `momentum' (space-time) coordinates $x^{i}$ together with
new `winding' coordinates $\tilde{x}_{i}$ and a covariant constraint that locally
eliminates half of the coordinates.
Originally accomplished for the bosonic string
\cite{Hull:2009mi,Hohm:2010jy,Hohm:2010pp}, more recently
it has been extended to heterotic \cite{Hohm:2011ex} and to type II superstrings
\cite{Hohm:2011zr,Hohm:2011dv}.
(For earlier and related work see 
\cite{Siegel:1993th,Tseytlin:1990nb,Kwak:2010ew,West:2010ev,Thompson:2011uw,Coimbra:2011nw,Berman:2010is,Jeon:2010rw,Copland:2011yh,Andriot:2011uh}.)
In particular, the type II double field theory provides a unified description of the
\textit{massless} type IIA and type IIB theories.
In this paper we will show that a minimal extension of this theory exists that
also contains \textit{massive} deformations \cite{Romans:1985tz}. 
(For earlier results on massive type IIA
within `duality-covariant' frameworks see \cite{Schnakenburg:2002xx,Kleinschmidt:2004dy,Henneaux:2008nr}.)

The massive extension of type IIA supergravity due to Romans \cite{Romans:1985tz}
can be motivated as follows. If one introduces for each RR $p$-form the dual $8-p$ form,
type IIA contains all odd forms with $p=1,\ldots,7$. We can also introduce a
9-form potential, but imposing the standard field equations sets  
its field strength $F^{(10)}=dC^{(9)}$ to a constant 
and so a 9-form carries no propagating degrees of freedom.
We can think of massive type IIA as obtained
by choosing this integration constant to be non-zero and equal to the mass parameter $m$.
In the resulting theory, $m$ enters
as a cosmological constant and deforms the gauge transformations corresponding
to the NS-NS $b$-field such that the RR 1-form transforms with a St\"uckelberg
shift symmetry. It does not admit a maximally symmetric vacuum, but its most symmetric solution
is the D8 brane solution that features 9-dimensional Poincar\'e invariance \cite{Bergshoeff:1996ui}.

The type II double field theory of \cite{Hohm:2011zr,Hohm:2011dv} is  formulated in terms
of a Majorana-Weyl spinor of the `T-duality group' $O(10,10)$. This  spinor representation is
isomorphic to the set of all even or odd forms, depending on the chirality of the spinor, and so
for type IIA the theory contains already a 9-form potential. However, the duality
relations
 \be
  * \widehat{F}^{(10)} \ = \ -  \widehat{F}^{(0)} \ \equiv \ 0
 \ee
imply that its field strength is zero, because there is no non-trivial
$\widehat{F}^{(0)}$ due to the absence of `$(-1)$--form' potentials, and therefore
the 9-form is on-shell determined to be pure gauge.
Formally, one may introduce a $(-1)$--form potential  $C^{(-1)}$ and then set $m=F^{(0)}=dC^{(-1)}$,
as has been done in \cite{Lavrinenko:1999xi}, but so far it has been unclear how to find
a mathematically satisfactory interpretation of such objects. In this note we will show
that a non-trivial 0-form field strength (and thus a
mass parameter) is naturally included in the type II double field theory by assuming
that the RR 1-form depends linearly on the winding coordinates,
 \be\label{ansatz}
  C^{(1)}(x,\tilde{x}) \ = \  C_{i}(x)dx^{i}+m\tilde{x}_{1} dx^{1}\;,
 \ee
where $C_{i}$ and all other fields depend only on the 10-dimensional coordinates.
We will see that the second term in (\ref{ansatz}) effectively acts as a $(-1)$--form and that
the double field theory reduces precisely to massive type IIA.

It should be stressed that the consistency of the ansatz (\ref{ansatz}) is non-trivial, in fact,
even unexpected, for
the double field theory formulation requires the constraint that
  \be\label{ODDconstr}
   \partial^{M}\partial_{M}A \ = \ \eta^{MN}\partial_{M}\partial_{N}A \ = \ 0\;, \qquad
   \partial^{M}A\,\partial_{M}B \ = \ 0\;, \qquad
   \eta^{MN} \ = \  \begin{pmatrix}
    0&1 \\1&0 \end{pmatrix}\,,
 \ee
for all fields and gauge parameters  $A, B$. 
Here, $M, N,\ldots$ are fundamental $O(10,10)$ indices, with invariant metric
$\eta_{MN}$, and the derivative $\partial_{M}=(\tilde{\partial}^{i},\partial_{i})$ combines
the partial momentum and winding derivatives.
This constraint is necessary for gauge invariance of the action and closure of the
gauge algebra. In its \textit{weak} form, which requires
$\partial^{M}\partial_{M}=2\tilde{\partial}^{i}\partial_{i}$ to annihilate
all fields and parameters, it is a direct consequence of the level matching condition of closed
string theory, and it allows for field configurations such as (\ref{ansatz}) that
depend locally both on $x$ and $\tilde{x}$. The double field theory constructions completed
so far, however, impose the stronger form that requires also all products of fields and parameters to be
annihilated by $\partial^{M}\partial_{M}$, corresponding to the second equation in  (\ref{ODDconstr}).
In this form the constraint implies that locally all fields depend only on half of the coordinates, 
and so (\ref{ansatz}) violates the strong constraint. Remarkably, as we will show here,
the gauge transformations can be reformulated on the RR fields so that 
the strong constraint can be relaxed. It cannot be relaxed to the weak constraint as formulated 
above, but it is sufficient for the ansatz (\ref{ansatz}) to be consistent. 
In particular, this formulation guarantees that
in the action and gauge transformations the linear $\tilde{x}$ dependence drops out,
such that the resulting theory has a conventional 10-dimensional interpretation.

This paper is organized as follows. In sec.~2 we briefly review the type II double
field theory of \cite{Hohm:2011zr,Hohm:2011dv}, and we give a rewriting of the gauge transformations
that requires only a weaker constraint in a sense to be made precise. 
In sec.~3 we evaluate the double field theory for
(\ref{ansatz}) and show that it reduces to the democratic formulation of massive type IIA.
We then discuss T-duality for massive type II by evaluating the double field
theory for a coordinate dependence in which one of the space-time coordinates
is interchanged with a winding coordinate, corresponding to a single T-duality inversion.
In the massless case this maps type IIA to type IIB, in agreement with their equivalence
upon compactification on a circle. In the massive case, however, it maps type IIA into
a non-covariant `massive' version of type IIB. In this formulation, type IIB 
has only manifest 9-dimensional covariance but is fully diffeomorphism invariant, with the 10th
diffeomorphism being deformed by the mass parameter.

\section{Type II double field theory}\setcounter{equation}{0}

\subsection{Massless theory}
We start by reviewing the double field theory formulation of massless type II theories.
The NS-NS fields are the space-time metric $g_{ij}$, the Kalb-Ramond 2-form $b_{ij}$ and
the dilaton $\phi$. The metric and $b$-field are encoded in the generalized metric
 \be\label{firstH}
  {\cal H}_{MN} \ = \  \begin{pmatrix}    g^{ij} & -g^{ik}b_{kj}\\[0.5ex]
  b_{ik}g^{kj} & g_{ij}-b_{ik}g^{kl}b_{lj}\end{pmatrix} \;,
 \ee
which is an element of $SO(10,10)$, satisfying the constraint ${\cal H}^{T}={\cal H}$ and
transforming covariantly under  $O(10,10)$ according to its index structure.
Moreover, we introduce an $O(10,10)$ invariant dilaton density by $e^{-2d}=\sqrt{g}e^{-2\phi}$,
where $g=| \det{g_{ij}}|$. We can think of ${\cal H}$ as the fundamental field and view
(\ref{firstH}) as a particular parametrization, but in
\cite{Hohm:2011zr} we argued that we should rather view the fundamental field as
an hermitian element of the two-fold covering group Spin$(10,10)$.
Denoting this field by $\fancys$, we require
 \be\label{basicS}
   \fancys \ = \ \fancys^{\dagger}\;, \qquad \fancys \ \in \ \text{Spin}(10,10)\;,
 \ee
and \textit{define} the generalized metric via the group homomorphism
$\rho:\;\,$Spin$(10,10)\rightarrow SO(10,10)$ as ${\cal H}=\rho(\fancys)$.
With (\ref{basicS}) we infer ${\cal H}^{T}=\rho(\fancys^{\dagger})=\rho(\fancys)={\cal H}$,
and thus, as required, ${\cal H}$ is a symmetric group element which generally can be
parametrized as in (\ref{firstH}).

In the democratic formulation to be employed here,
the RR fields consist of differential forms of all odd (even) degrees for type IIA (IIB) and
can be encoded in a Majorana-Weyl spinor $\chi$ of $O(10,10)$. In order to see this,
let us fix our conventions for Spin$(10,10)$. The Clifford algebra is given by
 \be\label{Clifford}
  \big\{ \Gamma^{M},\Gamma^{N}\big\} \ = \ 2\eta^{MN}\,{\bf 1}\;.
 \ee
Due to the off-diagonal form of the $O(10,10)$ metric in (\ref{ODDconstr}),
this algebra implies that
   \be \label{defgamma}
   \psi_{i} \ \equiv \ \frac{1}{\sqrt{2}}\,\Gamma_{i}\;, \quad
   \psi^{i} \ \equiv \ \frac{1}{\sqrt{2}}\,\Gamma^{i}\;,
 \ee
with $(\psi_i)^{\dagger}=\psi^i$, satisfy the canonical anti-commutation relations of fermionic raising
and lowering operators $\psi^i$ and $\psi_i$, respectively,
 \be \label{deffermosc}
  \{ \psi_i ,  \psi^j\} \ = \ \delta_{i}{}^{j}\;, \qquad
  \{\psi_i,\psi_j\} \ = \  0\;, \qquad \{\psi^i,\psi^j\} \ = \ 0\;.
 \ee
Thus, the spinor representation can be constructed by introducing
a Clifford vacuum $\ket{0}$, satisfying $\psi_i\ket{0}=0$ for all $i$, and acting
with the raising operators $\psi^{i}$. A spinor is then given by a general state
 \be \label{genstate}
  \chi \ = \ \sum_{p=0}^{10}\frac{1}{p!}\,C_{i_1\ldots i_p}\,\psi^{i_1}\ldots\psi^{i_p}\ket{0}\;,
 \ee
with coefficients that are fully antisymmetric tensors and that will be
identified with the RR $p$-forms $C^{(p)}$.  A spinor satisfying a chirality condition consists either of
only odd or only even forms, and in the following we assume that $\chi$ has a fixed chirality.
The Dirac operator
 \be\label{def-dir-op}
  \slashed\partial \ \equiv \  {1\over \sqrt{2}} \, \Gamma^M \partial_M
  \ = \   \psi^i\partial_i+\psi_i\tilde{\partial}^{i} \,,
 \ee
acts naturally on spinors and can be viewed as the $O(10,10)$ invariant extension
of the exterior derivative. First, for $\tilde{\partial}=0$ it only acts via differentiating
with respect to $x^{i}$ and increasing the form degree by one. Second, it squares to
zero,
\be\label{zerosquare}
  \slashed{\partial}^2 \ = \
  \frac{1}{2}\Gamma^{M}\Gamma^{N}\partial_{M}\partial_{N} \ = \
  \frac{1}{2}\eta^{MN}\partial_{M}\partial_{N} \ = \ 0\,,
 \ee
using (\ref{Clifford}) and the constraint (\ref{ODDconstr}).

The type II double field theory, whose independent fields are $\fancys$, $\chi$ and $d$,
is defined by the action
\be
\label{totaction}
 S \ = \  \int d^{10}x\, d^{10}\tilde x\, \Bigl(  e^{-2d}\, {\cal R} ({\cal H} , d)  + \frac{1}{4}
 (\slashed{\partial}{\chi})^\dagger \;
 \fancys \; \slashed\partial\chi\,\Bigr) \,,
\ee
supplemented by the self-duality constraint
 \be\label{dualityintro}
 \slashed\partial\chi \ = \ -{\cal K}\,\slashed\partial\chi\,, \quad
 {\cal K} \ \equiv \ C^{-1} \fancys\, ,
\ee
where $C$ denotess the charge conjugation matrix of Spin$(10,10)$. For the definition
of the $O(10,10)$
invariant scalar ${\cal R}({\cal H} , d)$ see eq.~(4.24) in \cite{Hohm:2010pp}.

Let us next review the symmetries of this theory. It has a global T-duality invariance under
Spin$^{+}(10,10)$, which is the component of the group connected to the identity. It is gauge invariant
under
 \be\label{abelian}
  \delta_{\lambda} \chi \ = \ \slashed{\partial} \lambda\,,
 \ee
where $\lambda$ is a spinor of Spin$(10,10)$ with a fixed chirality opposite to the one
of $\chi$, such that odd (even) $p$-forms transform with even (odd) $(p-1)$-form parameters.
This symmetry leaves (\ref{totaction}) and
(\ref{dualityintro}) manifestly invariant because of $\slashed{\partial}^2=0$.
The double field theory is also invariant under a `generalized diffeomorphism' symmetry
spanned by a parameter $\xi^{M}=(\tilde{\xi}_{i},\xi^{i})$ that combines the usual diffeomorphism
parameter $\xi^i$ and the $b$-field gauge parameter $\tilde{\xi}_{i}$ into a fundamental
$O(10,10)$ vector.
This gauge symmetry acts on the fundamental
fields as
\be\label{newxiM}
 \begin{split}
   \delta_{\xi}\big(e^{-2d}\big) \ &= \ \partial_{M}\big(\xi^{M}e^{-2d} \big)\;, \\
   \delta_{\xi} {\cal K} \ &= \ \xi^M \partial_M  {\cal K}
   + {1 \over 2} \big[ \Gamma^{MN},   {\cal K} \, \big]  \partial_{M} \xi_{N} \; , \\
  \delta_\xi \chi \ &= \ \xi^M \partial_M  \chi +  {1\over 2}   \partial_M \xi_N \Gamma^M
  \Gamma^N \chi\;, 
 \end{split}
 \ee
where $\Gamma^{MN}=\tfrac{1}{2}[\Gamma^{M},\Gamma^{N}]$,
 and we have written the gauge variation of $\fancys$ in terms of ${\cal K}$ defined
in (\ref{dualityintro}). The gauge invariance under $\xi^{M}$ transformations is non-trivial
and requires the strong constraint (\ref{ODDconstr}). We stress that, in contrast,
the identity (\ref{zerosquare}) and thus the invariance under $\lambda$ transformations
requires only the weak constraint, i.e., that $\partial^{M}\partial_{M}$ annihilates $\chi$
and its gauge parameter, but not necessarily their products, which will be used below.

Next we discuss the double field theory evaluated for  fields
depending only on $x^{i}$, i.e., setting the winding derivatives to zero, $\tilde{\partial}^{i}=0$.
In order to compare the resulting theory with the conventional formulation, we need
to choose a parametrization of $\fancys$ in terms of $g$ and $b$, as for ${\cal H}$ in
(\ref{firstH}). We do so by decomposing (\ref{firstH})
  \be
   {\cal H} \ = \  \begin{pmatrix} 1 &   0 \\ b & 1 \end{pmatrix}
   \begin{pmatrix} g^{-1} &   0 \\ 0 & g \end{pmatrix}
   \begin{pmatrix} 1 &   -b \\ 0 & 1 \end{pmatrix} \ \equiv \
   h_b^T\,h_{g^{-1}}\,h_{b} \; ,
 \ee
and then introducing spin representatives for each of the $SO(10,10)$ elements
on the right-hand side. We then set
\be
\label{definitionSH}
\phantom{\Biggl(}
\fancys \ = \
S_{\cal H} \ \equiv  \ S_b^{\dagger}\,  S_g^{-1}\, S_{b} \ = \  e^{\frac{1}{2}b_{ij}\psi_{i}\psi_{j}}\, S_{g}^{-1}\,
e^{ -\frac{1}{2}b_{ij}\psi^{i}\psi^{j}}
\;,
\ee
where we introduced the spin representative $S_{b}$ of $h_b$ 
in terms of fermionic oscillators \cite{Hohm:2011zr}. 
The explicit form of the spin representative $S_{g}$ of $h_{g}$ 
can be found in \cite{Hohm:2011zr}. By construction, $\fancys=S_{\cal H}$
projects under the group homomorphism $\rho$ to ${\cal H}$ in $SO(10,10)$, and it has been shown in \cite{Hohm:2010jy,Hohm:2010pp}
that the first term in (\ref{totaction}) reduces to the standard low-energy action for the
NS-NS sector. 

We turn now to the RR sector.
For $\tilde{\partial}^i=0$ the Dirac operator $\slashed{\partial}$ acts like the exterior
derivative and so $\slashed{\partial}\chi$ reduces to the conventional
field strengths of the RR potentials in (\ref{genstate}),
 \be\label{fieldstrengths}
    \slashed{\partial}\chi\, \Big|_{\tilde{\partial}=0} \ = \
   \sum_{p}\frac{1}{p!}\partial_{j}C_{i_1\ldots i_{p}}\,\psi^{j}\psi^{i_1}\cdots \psi^{i_{p}}\ket{0} \ = \
    \sum_{p}\frac{1}{(p+1)!}F_{i_1\ldots i_{p+1}}\,\psi^{i_1}\cdots \psi^{i_{p+1}}\ket{0}\;,
 \ee
where we introduced in the last equation the components of the $(p+1)$--form
field strength,
  \be
  \label{fsp}
  F^{(p+1)} \ = \ d C^{(p)} \,.
 \ee
In the action (\ref{totaction}) the exponentials of $b$ in (\ref{definitionSH}) lead to
the modified field strength  \cite{Hohm:2011zr,Hohm:2011dv}
 \be
  \label{ayvg}
  \widehat F \ = \ e^{-b^{(2)}} \wedge F \ = \
  e^{-b^{(2)}} \wedge dC  \;,
\ee
where we use a notation that combines all $p$-forms into a formal sum. Without
repeating the derivation, we recall that the $S_{g}$ factor in (\ref{definitionSH}) effectively
arranges the proper contractions with the space-time metric $g$ in the action, so that
the RR part of (\ref{totaction}) reads \cite{Hohm:2011zr,Hohm:2011dv}
 \be
 \label{action-reduction}
  S_{\rm RR} \ = \ -\frac{1}{4}\int d^{10}x\,\sqrt{g}\,\sum_{p}\,\frac{1}{p!}\,g^{i_1 j_1}\cdots g^{i_p j_p}\,
  \widehat{F}_{i_1\ldots i_p}\widehat{F}_{j_1\ldots j_p}
  \ = \ \frac{1}{4} \sum_{p}\int\, \widehat{F}^{(p)}\wedge\, *\widehat{F}^{(p)}\;.
 \ee
Similarly, the self-duality constraint (\ref{dualityintro}) reduces to the conventional
duality relations
 \be\label{d-rel10}
   \widehat{F}^{(p)} \ = \ - (-1)^{{1\over 2} p(p+1)} *\widehat{F}^{(10-p)} \; .
\ee
Thus, for the RR sector the double field theory action reduces to the sum of kinetic terms for all $p$-forms,
which is supplemented by duality relations.  This is the `democratic formulation' originally
introduced in \cite{Fukuma:1999jt}, and which is on-shell equivalent to the conventional
formulation of type IIA for odd forms or type IIB for even forms.

Let us finally review the gauge transformations for $\tilde{\partial}^i=0$. The gauge
variations (\ref{abelian}), with parameter
 \be
  \lambda \ = \ \sum_{p}\frac{1}{p!}\lambda_{i_1\ldots i_p}\,\psi^{i_1}\cdots\psi^{i_p}\ket{0}\;,
 \ee
reduce to
 \be
  \delta_{\lambda}\chi \ = \  \psi^{j}\partial_{j}\lambda \ = \
  \sum_{p}\frac{1}{(p-1)!}\partial_{[i_1}\lambda_{i_2\ldots i_p]}\,\psi^{i_1}\cdots\psi^{i_p}\ket{0}\;,
 \ee
which amounts to the standard abelian gauge symmetry of the RR $p$-forms,
  \be\label{pformlangCtr}
\delta_\lambda C \ = \ d \lambda \,.
\ee
The gauge symmetry (\ref{newxiM}) parameterized by $\xi^{M}=(\tilde{\xi}_{i},\xi^{i})$ gives for $\xi^{i}$
 \be
  \delta_{\xi}\chi \ = \ \left(\xi^{j}\partial_{j}+\partial_{j}\xi^{k}\,\psi^{j}\psi_{k}\right)
  \sum_{p}\,\frac{1}{p!}\,C_{i_1\ldots i_p}\,\psi^{i_1}\cdots \psi^{i_p}\ket{0}\;,
 \ee
which by use of the oscillator algebra (\ref{deffermosc}) implies
 \be\label{formdiff}
  \delta_{\xi}C_{i_1\ldots i_p} \ = \ \xi^{j}\partial_{j}C_{i_1\ldots i_p}
  +p\, \partial_{[i_1}\xi^{j}\,C_{|j|i_2\ldots i_p]} \ \equiv \ {\cal L}_{\xi}C_{i_1\ldots i_p}\;.
 \ee
This is the conventional diffeomorphism symmetry, acting infinitesimally via the Lie derivative.
From (\ref{newxiM}) it also follows that the $C^{(p)}$ transform non-trivially under the $b$-field
gauge parameter $\tilde{\xi}_{i}$,
 \be
   \delta_{\tilde{\xi}}\chi \ = \ \partial_{k}\tilde{\xi}_{l}\,\psi^{k}\psi^{l}\chi \ = \
  \sum_{p}\frac{1}{p!}\partial_{[i_1}\tilde{\xi}^{}_{i_2}\,C_{i_3\ldots i_{p+2}]}
  \psi^{i_1}\cdots \psi^{i_{p+2}}\ket{0} \;,
\ee
which implies
  \be
 \label{RRformbgauge2}
 \delta_{\tilde{\xi}} C \ = \  d\tilde \xi \wedge C \,.
 \ee
This means that the $C^{(p)}$ are redefinitions by the $b$-field of the more conventional RR fields,
which are invariant under $\tilde{\xi}_i$. These are exactly the expected gauge symmetries for
massless type II theories.

\subsection{Reformulation of gauge symmetries}
The gauge invariance of the type II double field theory requires for $\xi^{M}$ transformations
the strong form of the constraint (\ref{ODDconstr}), but for $\lambda$ transformations
only the weak constraint. Here, we will perform a change of basis for the gauge parameters
such that, for the RR sector, the $\xi^{M}$ transformations are consistent with a weaker form 
of the constraint.

We start by rewriting the $\xi^{M}$ gauge transformation of $\chi$ as follows
 \be
 \begin{split}
  \delta_{\xi}\chi \ &= \ \xi^{M}\partial_{M}\chi+\frac{1}{2}\partial_{M}\xi_{N}\Gamma^{M}\Gamma^{N}\chi\\
  \ &= \ \xi^{M}\partial_{M}\chi-\frac{1}{2}\Gamma^{M}\Gamma^{N}\xi_{N}\partial_{M}\chi
  +\frac{1}{2}\Gamma^{M}\partial_{M}\left(\xi_{N}\Gamma^{N}\chi\right)\;.
 \end{split}
 \ee
The last term is of the form of a field-dependent $\lambda$ gauge transformation $\slashed{\partial}\lambda$ 
and can therefore be ignored.
We then use the Clifford algebra in the second term,
 \be
  \delta_{\xi}\chi \ = \  \xi^{M}\partial_{M}\chi -\frac{1}{2}\big( 2\eta^{MN}-\Gamma^{N}\Gamma^{M}\big)
  \xi_{N}\partial_{M}\chi \ = \ \frac{1}{2}\Gamma^{N}\Gamma^{M}\xi_{N}\partial_{M}\chi\;.
 \ee
Using the `slash' notation (\ref{def-dir-op}),
 \be
  \slashed{\partial} \ = \ \frac{1}{\sqrt{2}}\Gamma^{M}\partial_{M}\;, \qquad
  \slashed{\xi}  \ = \ \frac{1}{\sqrt{2}}\Gamma^{M}\xi_{M}\;,
 \ee
we finally get
 \be\label{finalgauge}
  \delta_{\xi}\chi \ = \ \slashed{\xi}\,\slashed{\partial}\chi\;,
 \ee
which is the form of the $\xi^{M}$ gauge transformations we will use from now on.

We will show that, starting from (\ref{finalgauge}),  
gauge invariance of the RR action and closure of the gauge algebra uses only the constraint
 \be\label{weakconstr}
  \eta^{MN}\partial_{M}\partial_{N}A \ = \ \partial^{M}\partial_{M}A\ = \ 0\;, \qquad
  A \ = \ \big\{\chi,\lambda,\xi^{M}\big\}\;.
 \ee 
In particular, for this computation 
we do not need to use $\partial^{M}A\partial_{M}B=0$.  
This observation does not imply, however, that the RR sector is `weakly constrained' in the sense 
that fields but not their products need to satisfy the constraint. 
In fact, (\ref{finalgauge}) is not a consistent transformation rule 
assuming that $\chi$ and $\xi$ are weakly constrained.  
Before discussing this in more detail,  
we investigate some consequences of the form (\ref{finalgauge}) of the 
gauge transformations.

The original gauge transformations have the property
that a gauge parameter of the form $\xi^{M}=\partial^{M}\Theta$ is `trivial' in that it
generates no gauge transformation.
After the above redefinition, this statement is modified.  We compute
 \be\label{step01}
  \delta_{\partial \Theta}\chi \ = \
  \frac{1}{2}\Gamma^{N}\Gamma^{M}\partial_{N}\Theta\,\partial_{M}\chi \ = \
  \frac{1}{2}\Gamma^{N}\partial_{N}\big(\Theta\,\Gamma^{M}\partial_{M}\chi\big)\;,
 \ee
assuming only the weaker form (\ref{weakconstr}) of the constraint. 
Thus, the gauge variation (\ref{step01}) takes the form of a field-dependent $\lambda$ gauge
transformation,
 \be
   \delta_{\partial\Theta}\chi \ = \    \slashed{\partial}\lambda\;, \qquad
   \lambda \ = \ \Theta\, \slashed{\partial}\chi \;.
 \ee
Therefore, the statement that $\xi^{M}=\partial^{M}\Theta$ leads to a trivial gauge transformation
leaving the fields invariant has to be relaxed to the statement that it leaves the fields invariant
\textit{up to a $\lambda$ gauge transformation}, but it has the advantage that in this weaker form
only the constraint (\ref{weakconstr}) is required.

We compute next the gauge variation of $\slashed{\partial}\chi$ under (\ref{finalgauge}),
which is needed in order to verify gauge invariance,
 \be
 \begin{split}
  \delta_{\xi}\big(\slashed{\partial}\chi\big) \ &= \ \slashed{\partial}\big(\slashed{\xi}\,\slashed{\partial}\chi\big)
  \ = \  \slashed{\partial}\slashed{\xi}\,\slashed{\partial}\chi +  \frac{1}{2\sqrt{2}}
  \Gamma^{M}\Gamma^{N}\Gamma^{P}\xi_{N}\partial_{M}\partial_{P}\chi    \\
  \ &= \  \slashed{\partial}\slashed{\xi}\,\slashed{\partial}\chi +  \frac{1}{2\sqrt{2}}
  \big(2\eta^{MN}-\Gamma^{N}\Gamma^{M}\big)\Gamma^{P} \xi_{N}\partial_{M}\partial_{P}\chi  \;.
 \end{split}
 \ee
The last term contains $\Gamma^{(M}\Gamma^{P)}=\eta^{MP}$ and therefore vanishes
by (\ref{weakconstr}), while the second term reduces to
$\xi^{M}\partial_{M}\slashed{\partial}\chi$.  In total we have
 \be\label{fieldstrvar}
  \delta_{\xi}\big(\slashed{\partial}\chi\big) \ = \ \xi^{M}\partial_{M}\slashed{\partial}\chi
  + \slashed{\partial}\slashed{\xi}\,\slashed{\partial}\chi \;.
 \ee
This result agrees with the variation under the original form of the gauge transformations
determined in \cite{Hohm:2011zr}
(as it should be, because the modification is  a $\lambda$ gauge transformation that leaves
$\slashed{\partial}\chi$ invariant), but in the original derivation the strong constraint was used.
As the proof of gauge invariance of the action and the self-duality constraint given in
\cite{Hohm:2011zr} requires only
the transformation rule (\ref{fieldstrvar}),
we conclude that this proof uses only the weaker constraint (\ref{weakconstr}).

Let us verify that also closure of the gauge transformations on $\chi$ requires only this 
weaker constraint. First, for the modified form of the gauge transformations
there is no non-vanishing commutator
between $\lambda$ and $\xi$ gauge transformations because $\slashed{\partial}\chi$
is $\lambda$-invariant. Thus, it remains to verify closure of the $\xi^{M}$ transformations, for
which we find
  \be\label{xiclosure}
   \big[ \delta_{\xi^1}, \delta_{\xi^2}\big]\chi
      \ = \ \delta_{\xi^{12}}\chi+\delta_{\lambda}\chi \;.
 \ee
Here,
  \be\label{xparam}
  \xi_{12}^{M} \ = \ \big[\xi_{2},\xi_{1}\big]_{\rm C}^{M} \ = \
  \xi_{2}^{N}\partial_N \xi_{1}^{M}-\frac{1}{2}\xi_{2N}\partial^{M}\xi_1^{N}-(1\leftrightarrow 2)\;,
 \ee
which is given by the `C-bracket' that characterizes the closure of $\xi^{M}$
transformations on the NS-NS fields  \cite{Hull:2009mi,Hohm:2010jy}, and
  \be\label{lambdaparam}
  \lambda \ = \
  -\frac{1}{2}\big(\slashed{\xi}_2\slashed{\xi}_1-\slashed{\xi}_1\slashed{\xi}_2\big)
  \slashed{\partial}\chi\;.
 \ee
The verification of (\ref{xiclosure}) is a straightforward though somewhat tedious 
exercise in gamma matrix algebra, which we defer to the appendix.  
The computation makes repeated use of the
constraint (\ref{ODDconstr}), but only in its relaxed form (\ref{weakconstr}). Thus, on the RR field $\chi$
all gauge symmetries close using only this weaker constraint.

We close this section by computing the form of these redefined  $\xi^{M}$ gauge
transformations (\ref{finalgauge})
for $\tilde{\partial}^i=0$. For the diffeomorphism parameter $\xi^{i}$ we find
 \be
\delta_{\xi}\chi \ = \  \xi^{i}\psi_{i}\psi^{j} \partial_{j}\chi \ = \
\sum_{p } \frac{1}{p!}  {\xi}^i \partial_{j}
C_{i_1 \cdots i_p}  \psi_i \psi^j  \psi^{i_1} \cdots \psi^{i_p} \ket{0}  \; .
\ee
Using the oscillator algebra (\ref{deffermosc}) to simplify this, we obtain
 \be\label{improve}
  \delta_{\xi} C_{i_1 \cdots i_p} \ = \
  (p+1) \xi^{j} \partial_{[j} C_{i_1 \cdots i_p]} \ = \  \xi^{j} F_{j i_1 \cdots i_p} \; .
\ee
For the $b$-field gauge parameter $\tilde{\xi}_{i}$ one obtains
 \be
  \delta_{\tilde{\xi}}\chi \ = \ \tilde{\xi}_{i}\psi^{i}\psi^{j}\partial_{j}\chi
  \ = \ \sum_p\frac{1}{p!}\,\tilde{\xi}_{i}\partial_{j}
  C_{i_1\ldots i_p}\psi^{i}\psi^{j}\psi^{i_1}\cdots \psi^{i_p}\ket{0}\;,
 \ee
from which we read off
 \be\label{newxitilde}
   \delta_{\tilde{\xi}}C \ =  \ \tilde{\xi}\wedge F\;.
 \ee

The diffeomorphism symmetry in the form (\ref{improve}) is sometimes referred to as
`improved diffeomorphisms'. They can be introduced for any $p$-form gauge field
by adding to the familiar 
diffeomorphism symmetry (\ref{formdiff}) a field-dependent gauge transformation with
$(p-1)$-form parameter
 \be
  \lambda_{i_1\ldots i_{p-1}} \ = \ -\xi^{j}C_{ji_1\ldots i_{p-1}}\;.
 \ee
Similarly, (\ref{newxitilde}) is obtained from the original $\tilde{\xi}$ transformation
(\ref{RRformbgauge2}) by adding an abelian gauge transformation with parameter
$\lambda=-\tilde{\xi}\wedge C$.
Thus, the redefinition of the gauge transformations leading to (\ref{finalgauge})
is precisely the double field theory analogue of the improved diffeomorphisms in conventional
gauge theories. In this form the gauge field appears only under a derivative,
which will be instrumental for the generalization we discuss next.

\section{Massive type II theories}\setcounter{equation}{0}

\subsection{Massive type IIA}
In the previous section we have seen that the proof of gauge invariance and closure of the 
gauge algebra uses only the weaker constraint (\ref{weakconstr}) for the RR sector. 
Naively, this would allow
for field configurations like 
 \be\label{chiexpand}
  \chi(x,\tilde{x}) \ = \ \chi_0(x)+\chi_1(\tilde{x})\;, 
 \ee
where $\chi_{0,1}$ are arbitrary functions of their arguments, 
and similarly for the gauge parameters. However, as mentioned above, there is a subtlety,  
because the gauge variations (\ref{finalgauge}) are not consistent 
assuming only the weak constraint.    
In fact, $\delta_{\xi}\chi$ on the left-hand side should satisfy the constraint, but with 
$\chi$ and $\xi$ being weakly constrained their product on the right-hand side in 
general does not satisfy the constraint. 
Rather, one should introduce a projector that restricts to the part  
satisfying the weak constraint \cite{Hull:2009mi}, while our computation above 
did not  keep track of these projectors. After the insertion of projectors, 
the gauge invariance of the action and closure of the gauge algebra does not follow from 
our computation (and is most likely not true). 
Moreover, the RR fields interact with the NS-NS sector that is still strongly constrained, and so 
it is presumably inconsistent to have a weakly constrained RR sector. 
Thus, a complete relaxation of the strong constraint must await 
a resolution of this problem for the NS-NS sector. However, if we 
only assume the function $\chi_1$ in (\ref{chiexpand}) to depend 
linearly on $\tilde{x}$, the resulting gauge variations and 
field equations are independent of $\tilde{x}$, and therefore the constraint is satisfied  
without insertion of projectors. (In particular, the energy-momentum tensor of the RR
fields depends only on $\slashed{\partial}\chi$ \cite{Hohm:2011zr} and is thereby   
independent of $\tilde{x}$.) An ansatz with linear $\tilde{x}$ dependence 
is therefore consistent, and we will investigate its consequences in what follows.

We will show that the type II double field theory defined by (\ref{totaction}) and
(\ref{dualityintro}) leads to massive type IIA if we assume that the RR spinor $\chi$
depends on the 10-dimensional space-time coordinates and, in its 1-form part, also 
linearly on a  winding coordinate. We thus write
 \be \label{mstate}
  \chi(x,\tilde{x}) \ = \ \Big(\sum_{p}\frac{1}{p!}\,C_{i_1\ldots i_p}(x)\,\psi^{i_1}\ldots\psi^{i_p}
  +m\tilde{x}_1\psi^1\Big)\ket{0}\;,
 \ee
where we assume that $\chi$ is of negative chirality such that the sum extends only over odd $p$.
Here we have singled out a particular (winding) coordinate direction, but we stress that this choice is
immaterial for the final result: we could have chosen any linear combination of the
$\tilde{x}_{i}$, which would merely amount to a rescaling of the mass parameter $m$.
Let us also note that it would be consistent to allow for a linear $\tilde{x}$ dependence
in other $p$-form parts, both in $\chi$ and in its gauge parameter $\lambda$.
We will comment on this more general case below.

Let us next evaluate the field strength $\slashed{\partial}\chi$  for  (\ref{mstate}).
In contrast to (\ref{fieldstrengths}), the term $\psi_{i}\tilde{\partial}^{i}$ in
$\slashed{\partial}$ acts now non-trivially,
 \be\label{mfield}
\begin{split}
\slashed{\partial}\chi & \ = \    \sum_{p}\frac{1}{p!}\partial_{j} C_{i_1\ldots i_{p}}\,\psi^{j}\psi^{i_1}\cdots \psi^{i_{p}}\ket{0} + \psi_j \tilde{\partial}^{j} ( m \tilde{x}_1) \psi^1 \ket{0} \\
& \ = \ \sum_{p}
\frac{1}{(p+1)!}(p+1)\partial_{[i_1} C_{i_2\ldots i_{p+1}]}\,
\psi^{i_1}\cdots \psi^{i_{p+1}}\ket{0}  + m \ket{0} \\
& \ \equiv \ \sum_{p}\frac{1}{(p+1)!}(F_{m})_{i_1\ldots i_{p+1}}\psi^{i_1}\cdots \psi^{i_{p+1}}\ket{0}\; ,
\end{split}
\ee
where we used the oscillator algebra (\ref{deffermosc}). We observe that the
non-trivial action of $\psi_{i}\tilde{\partial}^{i}$ leads to a reduction of the
form degree such that the `1-form potential' precisely leads to a non-vanishing 0-form field strength or, in
other words, that the $\tilde{x}$ dependent part acts effectively like a `$(-1)$-form'.
The $m$-deformed field strengths defined in the last line of (\ref{mfield})  then read
 \be\label{mfieldcomp}
  F_m^{(0)} \ = \  m \ , \qquad F_m^{(p+1)} \ = \ F^{(p+1)} \ = \ d C^{(p)} \;\; \quad
  \text{for\;\;\; $p \geq 0$ } \; .
 \ee
In the action the modified field strengths (\ref{ayvg}) enter, which are now deformed
according to (\ref{mfieldcomp}),
 \be\label{mdefF}
   \widehat{F}_m \ = \ e^{- b^{(2)}} \wedge (d C + m) \; .
\ee
This reads explicitly
\be\label{Fmcomp}
\begin{split}
\widehat{F}^{(0)}_m  & \ =  \   m \\  \widehat{F}^{(2)}_m & \ = \ F^{(2)} - mb^{(2)}   \\
\widehat{F}_m^{(4)} & \ = \ F^{(4)} - b^{(2)} \wedge F^{(2)} + \frac{1}{2} m b^{(2)} \wedge b^{(2)} \; ,
\quad {\rm etc.}
\end{split}
\ee
These are precisely the $m$-deformed field strengths appearing in massive type IIA,
see, e.g., \cite{Bergshoeff:1996ui}.

We turn now to the gauge symmetries acting on (\ref{mstate}), starting with the
$\lambda$-transformations  (\ref{abelian}). In analogy to (\ref{mstate}) it is
natural to allow here also for a linear $\tilde{x}$ dependence in the 0-form part
of $\lambda$, but such a contribution will be annihilated by $\slashed{\partial}$
due to $\psi_i\ket{0}=0$.  We note, however, that a linear 
$\tilde{x}$ dependence in the higher-form components of $\lambda$ can lead to 
a \textit{rigid} shift of the RR forms, which is trivially a global 
symmetry since all RR potentials appear under a derivative. 
We conclude that the $\lambda$ gauge transformations
are unchanged compared to the massless case (\ref{pformlangCtr}). 
The $\xi^{M}$ transformation (\ref{finalgauge}) evaluated for the diffeomorphism
parameter $\xi^{i}$ yields no new contribution since
 \be
  \delta_{\xi^i}\chi \, \Big|_{\partial_i=0} \ = \ \xi^{i}\psi_i\psi_j \tilde{\partial}^{j}\chi \ = \ 0\;,
 \ee
due to the action of two annihilation operators $\psi_i$  on (\ref{mstate}). Thus,
the diffeomorphism symmetry is given by (\ref{improve}), as for $m=0$.
Finally, the gauge transformation of the $b$-field gauge parameter
$\tilde{\xi}_i$ receives a  non-trivial modification,
 \be
  \delta_{\tilde{\xi}_i}\chi \, \Big|_{\partial_i=0} \ = \ \tilde{\xi}_{i}\psi^i \psi_j\tilde{\partial}^{j}\chi
  \ = \ m \tilde{\xi}_{i}\psi^i\psi_1\psi^1\ket{0} \ = \ m\tilde{\xi}_{i}\psi^i\ket{0}\;.
 \ee
Together with the gauge transformation (\ref{newxitilde}) for $m=0$ we thus obtain
\be\label{Stuckelberg}
 \delta_{\tilde{\xi}} C \ = \ \tilde{\xi} \wedge d C + m \tilde{\xi} \; .
\ee
Therefore, for $m\neq 0$ the RR 1-form $C^{(1)}$ transforms with a St\"uckelberg shift
symmetry under the $b$-field gauge transformations, which is precisely the expected result
for massive type IIA \cite{Bergshoeff:1996ui}.
The modified field strengths $\widehat{F}_m$ are manifestly invariant under
the $\lambda$ gauge transformations. The invariance under $\tilde{\xi}$ transformations
can be easily verified with $\delta_{\tilde{\xi}} b^{(2)} =  d \tilde{\xi}$,
\be
\begin{split}
\delta_{\tilde{\xi}}  \widehat{F}_m &\ = \ \delta_{\tilde{\xi}} \big( e^{- b^{(2)}} \wedge (dC + m) \big)
\ = \ - d \tilde{\xi} \wedge \widehat{F}_m + e^{- b^{(2)}} \wedge d \big(\tilde{\xi} \wedge d C + m \tilde{\xi} \big) \\
& \ = \  - d \tilde{\xi} \wedge \widehat{F}_m + e^{- b^{(2)}} \wedge d \tilde{\xi} \wedge (d C + m) \ = \ - d \tilde{\xi} \wedge \widehat{F}_m +  d \tilde{\xi} \wedge \widehat{F}_m \ = \ 0 \; .
\end{split}
\ee

\medskip
Let us now consider the double field theory action and duality relations  (\ref{totaction}) and
(\ref{dualityintro}), evaluated for (\ref{mstate}), and compare with the dynamics
of massive type IIA. As in (\ref{action-reduction}), the action reduces to the sum
of kinetic terms, but here for the modified field strengths  (\ref{mdefF}),
 \be\label{mdefkinetic}
  {\cal L}_{\rm RR} \ = \ \frac{1}{4}\,  \sum_{p = 0}^{10}\,
   \widehat{F}_m^{(p)} \wedge * \widehat{F}_m^{(p)} \ = \
  \frac{1}{4}\,  \sum_{p\geq 1}^{10} \widehat{F}_m^{(p)} \wedge * \widehat{F}_m^{(p)}
  +\frac{1}{4}m^2 *1 
  \, \; .
 \ee
The action contains now also the 0-form field strength, which contributes a cosmological
term proportional to $m^2$, as made explicit in the second equation.
Moreover, we can use the St\"uckelberg gauge symmetry (\ref{Stuckelberg})
with parameter $\tilde{\xi}$ to set $C^{(1)}=0$. From the second equation in
(\ref{Fmcomp}) we then infer that the kinetic term for $C^{(1)}$ reduces to a
mass term for the $b$-field. Thus, the $b$-field becomes massive by `eating' the RR 1-form.

The self-duality constraint (\ref{dualityintro}) reduces to the same duality relations
as in (\ref{d-rel10}), again with all field strengths being $m$-deformed,
 \be\label{d-rel10m}
   \widehat{F}_m^{(p)} \ = \ - (-1)^{{1\over 2} p(p+1)} *\widehat{F}_m^{(10-p)} \; .
\ee
This democratic formulation is equivalent to the conventional formulation of
massive type IIA. 
In the following we compare the two formulations in a little more detail.

The RR action of massive type IIA in the standard formulation is given by 
\cite{Romans:1985tz,Bergshoeff:1996ui}
 \be \label{standardRRaction}
\begin{split}
S_{\rm RR}
\ = \ &  \,\frac{1}{ 2} \int   \Bigl( \,
\widehat{F}_m^{(2)} \wedge *\widehat{F}_m^{(2)}  + \widehat{F}_m^{(4)} \wedge *\widehat{F}_m^{(4)}
 + m^2 *1 \,\Bigr)
  \\
&+   \frac{1}{ 2} \int \Bigl(  b^{(2)} (d C^{(3)})^2
 - (b^{(2)})^2 d C^{(1)} d C^{(3)} + \frac{1}{3} (b^{(2)})^{3} ( d C^{(1)} )^2 \\
&\qquad\qquad  + \frac{1}{3}m (b^{(2)})^{3}   d C^{(3)}
 - \frac{1}{4}m (b^{(2)})^4  d C^{(1)} + \frac{1}{20} m^2 (b^{(2)})^{5} \Bigr) \; ,
\end{split}
\ee
where for simplicity we have omitted
all wedge products between forms in the topological Chern-Simons terms $S_{\rm CS}$
in the second and third line.\footnote{This action differs from eq.~(2.8) of
\cite{Bergshoeff:1996ui} in certain numerical factors, which is due to different conventions regarding
differential forms. Moreover, there is a mismatch of a 
relative factor of $\tfrac{1}{2}$ between kinetic and Chern-Simons terms, but (\ref{standardRRaction}) 
is consistent with \cite{Romans:1985tz}.} 
We note in passing that this Chern-Simons action simplifies significantly if we formally
introduce a $(-1)$-form $C^{(-1)}$ and then define
 \be\label{Ahat}
  \widehat{A} \ = \ e^{-b^{(2)}} \wedge \big(C + C^{(-1)}\big) \; ,
\ee
where $C$ still represents the formal sum of all (odd) $p$-forms with $p\geq 1$.
The Chern-Simons
action can then simply be written as  
\be\label{simplestRRaction}
S_{\rm CS} \ = \  \frac{1}{2} \int b^{(2)} \wedge d \widehat{A}^{(3)} \wedge d \widehat{A}^{(3)} \; .
\ee
More precisely, expanding (\ref{simplestRRaction})
according to (\ref{Ahat}), the resulting action can be written, up to total derivatives, 
such that $C^{(-1)}$ enters only under 
an exterior derivative, and then setting $m=dC^{(-1)}$ reproduces precisely the 
Chern-Simons terms in (\ref{standardRRaction}). Formally, this drastic simplification can
be understood as a consequence of the $b$-field gauge transformations (\ref{Stuckelberg}),
which we rewrite here as 
 \be
  \delta_{\tilde{\xi}}C \ = \ \tilde{\xi}\wedge d\big(C+C^{(-1)}\big) \ = \ 
  d\tilde{\xi}\wedge \big(C+C^{(-1)}\big)-d\big(\tilde{\xi}\wedge\big(C+C^{(-1)}\big)\big)\;.
 \ee
The last term takes the form of a field-dependent $\lambda$ gauge transformation and 
can thus be ignored. The $\widehat{A}$ defined in (\ref{Ahat}) is then $\tilde{\xi}$
gauge invariant, 
 \be
  \delta_{\tilde{\xi}}\widehat{A} \ = \  -d\tilde{\xi} \wedge e^{-b^{(2)}}\wedge\big(C+C^{(-1)}\big)
  +e^{-b^{(2)}}\wedge\big(d\tilde{\xi}\wedge\big(C+C^{(-1)}\big)\big) \ = \ 0 \;,
 \ee
where we have taken $C^{(-1)}$ to be gauge invariant.  
From this we infer that (\ref{simplestRRaction}) is the only term invariant under $\tilde{\xi}$ gauge 
transformations (up to a boundary term). Note that we could have included the $(-1)$-form 
potential into the sum of all $p$-forms, in which case the gauge transformations 
would be formally as in the massless case.

We have verified the exact equivalence between the
equations of motion following from (\ref{standardRRaction}) and those
derived by varying (\ref{mdefkinetic}) and then supplementing them by the
duality relations (\ref{d-rel10m}). For the Einstein equations this is easy to see 
because the Chern-Simons terms that are present in the conventional formulation do
not contribute to the variation of the metric. The energy-momentum tensor 
then agrees for both formulations owing to the relative 
factor of $\tfrac{1}{2}$ between the kinetic terms in (\ref{mdefkinetic}) and (\ref{standardRRaction}), 
which compensates for the doubling of fields in the democratic formulation. 
For the field equations of the $p$-forms the on-shell equivalence is a consequence
of the Bianchi identities
\be
d \widehat{F}_m^{(p)} \ = \ - H^{(3)} \wedge  \widehat{F}_m^{(p-2)} \;,
\ee
following from (\ref{mdefF}). More precisely, the duality relations yield  
the second-order field equations as integrability conditions of $d^2=0$,  
including the required source terms originating from the 
Chern-Simons terms in the conventional formulation.  
Thus, the double field theory leads precisely
to massive type IIA.

\subsection{T-duality and massive type IIB}
We discuss now the double field theory evaluated for fields depending  
on coordinates that result from the 10-dimensional space-time coordinates $x^{i}$
by a T-duality inversion. 
The $O(10,10)$ 
invariance of the constraint (\ref{ODDconstr}) implies that fields  
resulting by an $O(10,10)$ transformation from fields depending only on the $x^{i}$ 
(thereby satisfying the constraint) 
also satisfy the constraint.  For instance, we may perform a single T-duality inversion 
in one direction, which exchanges a `momentum coordinate' $x^{i}$ with the 
corresponding `winding coordinate' $\tilde{x}_{i}$.  The double field theory 
evaluated for this field configuration then reduces to the T-dual theory. 
If it reduces to type IIA in one `T-duality frame',
it reduces to type IIB in the other frame, when expressed in the right T-dual field variables 
\cite{Hohm:2011zr}. The mapping of (massless) type IIA into 
type IIB under T-duality can therefore be discussed without reference to dimensional reduction, 
while in the usual approach this relation is inferred from the equivalence of type IIA and 
type IIB upon reduction on a circle \cite{Bergshoeff:1995as}.

Our task is now to see how this generalizes in the massive case. The usual point of view 
is as follows \cite{Bergshoeff:1996ui}. Massive type IIA reduced on a circle leads 
to a massive $N=2$ theory in nine dimensions, but there is no corresponding 
massive deformation of type IIB that could lead to the same nine-dimensional 
theory upon standard reduction.  Rather, to identify the proper T-duality rules  one has to perform 
a Scherk-Schwarz reduction \cite{Scherk:1979zr} of massless type IIB, which 
introduces a mass parameter and leads to the same massive $N=2$ theory 
in nine dimensions. In contrast, in double field theory the T-dual theory is 
identified without any dimensional reduction, as we discussed above, 
and so the puzzle arises what the 
T-dual to massive type IIA is if there is no massive type IIB in ten dimensions. 

In order to address this issue let us analyze the double field theory evaluated 
for fields in which one space-time coordinate, say $x^{10}$, is replaced 
by the corresponding winding coordinate. We split the coordinates 
as $x^{i}=(x^{\mu},x^{10})$, $\mu=1,\ldots,9$, 
and replace (\ref{mstate}) by the ansatz   
 \be
  \chi(x,\tilde{x}) \ = \ \Big(\sum_{p} \frac{1}{p!} C_{i_1\ldots i_p}(x^{\mu},\tilde{x}_{10})\,
  \psi^{i_1}\cdots \psi^{i_p}+m\tilde{x}_1\psi^1\Big)\ket{0}\;,
 \ee
where again the sum extends over all odd $p$. In the massless case  
the double field theory reduces to type IIB,  which can be made manifest by  
performing a field redefinition that takes the form of a T-duality inversion in the 
10th direction \cite{Hohm:2011zr}.\footnote{Here we assume that $x^{10}$ is a 
space-like direction, $g_{10,10}>0$. For T-dualities along time-like directions 
the dual theories are the so-called type II$^{*}$ theories \cite{Hohm:2011zr,Hohm:2011dv}, 
which have a reversed sign for the RR kinetic terms \cite{Hull:1998vg}. 
Similarly, the double field theory discussed here contains also a massive type IIA$^{*}$.} 
This T-duality transformation acts on the RR spinor via the 
spin representative $S_{10}=\psi^{10}+\psi_{10}$, i.e., we define 
 \be\label{trasndchi}
 \begin{split}
  \chi^{\prime} \ &= \ S_{10}\,\chi \ = \ 
  \Big( \sum_{p} \frac{1}{p!} C_{i_1\ldots i_p}^{\prime}\,\psi^{i_1}\cdots\, \psi^{i_p}
  +m\tilde{x}_1(\psi^{10}+\psi_{10})\psi^{1}\Big)\ket{0} \;, 
 \end{split}
 \ee
where in the first term we introduced redefined variables denoted by $C^{\prime}$. As 
$S_{10}$ is linear in the fermionic oscillators the sum extends now over all 
\textit{even} $p$. Specifically, one finds  (compare eq.~(6.41) in \cite{Hohm:2011dv}) 
\be\label{fform}
   C^{\prime}_{i_1\ldots i_p} \ = \   \left\{
  \begin{array}{l l}
   C_{\mu_2\ldots \mu_p} & \quad \text{if\;\; $i_1=10$,\; $i_2=\mu_2\,,\,\ldots$ $\,,i_p=\mu_p$}\\
   C_{10 \mu_1\ldots \mu_p}  & \quad \text{if\,\; $i_1=\mu_1\,,\, \ldots$ $\,,i_p=\mu_p$\,.}\\
  \end{array} \right.   
\ee
Thus, the dual field variables are obtained by adding or deleting the special index, thereby 
mapping odd forms into even forms, as required for the transition from type IIA to type IIB.
By performing this field redefinition (and renaming the coordinates) one infers   
that evaluating the theory for fields depending on $x^{\mu}$ and $\tilde{x}_{10}$ 
is equivalent to evaluating the theory for fields depending on $x^{i}$, but with 
the opposite chirality for the spinor, i.e., replacing odd forms by even forms. (See 
sec.~6.2 in \cite{Hohm:2011dv} for more details.) 
Now, in the massive case we have to take into account the second term in (\ref{trasndchi}), 
which reduces to $m\tilde{x}_1\psi^{10}\psi^{1}$.  
Thus, our task is to evaluate the double field theory for 
 \be\label{MIIBansatz}
  \chi(x,\tilde{x}) \ = \  \Big(\sum_{p} \frac{1}{p!} C_{i_1\ldots i_p}(x)\,
  \psi^{i_1}\cdots \psi^{i_p}+m\tilde{x}_1\psi^{10}\psi^{1}\Big)\ket{0}\;,
 \ee
dropping the primes from now on. In other words, we have to evaluate the double field theory 
for a field configuration in which the 2-form part depends now linearly on $\tilde{x}$, 
 \be\label{2Form}
  \big( \chi (x,\tilde{x})\big|_{\rm 2-form}\big)_{ij} \ = \  
  C^{}_{ij}(x)+2m\tilde{x}_1\,\delta_{[i}{}^{10}\,\delta_{j]}{}^{1}\;,
 \ee
with all other fields still depending only on the 10-dimensional space-time coordinates.

We start by computing the field strength
 \be
  F \ = \ \slashed{\partial}\chi  \ = \ F_{m=0} - 
  \psi_1\tilde{\partial}^{1}(m\tilde{x}_1)\psi^{1}\psi^{10}\ket{0} 
  \ = \  F_{m=0}- m\psi^{10}\ket{0}\;.
 \ee
Therefore, the field strength of the RR 0-form $C^{(0)}$ gets modified 
in the 10th component,
 \be\label{10thcompF}
  F^{(1)} \ = \ dC^{(0)}-mdx^{10}\;\qquad \Leftrightarrow\qquad \;
  F_{i} \ = \ \partial_{i}C^{(0)}-m\delta_{i}{}^{10}\;,
 \ee
while all other field strengths $F^{(p)}$, $p\neq 1$, remain unchanged.  
The `hatted' field strength (\ref{ayvg}) then receives corresponding modifications, 
 \be\label{IIBmdef}
  \widehat{F} \ = \ e^{-b^{(2)}}\wedge \big(dC-mdx^{10}\big)\;, 
 \ee
and thus in components 
 \be\label{FcompIIB}
  \widehat{F}^{(3)} \ = \ F^{(3)}-b^{(2)}\wedge dC^{(0)}+mb^{(2)}\wedge dx^{10}\;, \quad {\rm etc.}
 \ee
The dynamics is described by the same action (\ref{action-reduction}) and 
duality relations (\ref{d-rel10}) as before, but with all field strengths replaced by their  
$m$-deformed version (\ref{IIBmdef}). 

This theory breaks manifest 10-dimensional 
covariance in that the 10th coordinate is treated on a different footing 
in  (\ref{10thcompF}). We observe, however, that this theory can be obtained from standard 
(covariant) type IIB by performing the redefinition 
 \be\label{IIBredf}
  C^{(0)}\;\rightarrow\; C^{(0)}-mx^{10}\;,
 \ee
as is apparent from (\ref{10thcompF}). Thus, the `deformation' induced by the $m$-dependent 
2-form contribution in (\ref{MIIBansatz}) can be absorbed into a redefinition of the lower 
RR form $C^{(0)}$, and therefore the obtained theory is nothing but standard type IIB after a somewhat
peculiar (non-covariant) redefinition. For this reason we do not introduce a new symbol 
for the `deformed' field strengths.

In order to understand the consequences of the non-covariance 
let us inspect the gauge symmetries.
As above, the $\lambda$ gauge transformations are unchanged compared to the massless case.  
The gauge transformations (\ref{finalgauge}) 
parametrized by $\xi^{M}$ applied to (\ref{MIIBansatz}) give 
 \be
 \begin{split}
  \delta_{\xi}\chi \ &= \  \big(\psi^i\tilde{\xi}_i+\psi_{i}\xi^{i}\big)\slashed{\partial}\chi
  \ = \ \delta_{\xi}\chi \Big|_{m=0} -m (\tilde{\xi}_{i}\psi^{i}\psi^{10}+\xi^{i}\psi_{i}\psi^{10})\ket{0} \\
  \ &= \ 
 \delta_{\xi}\chi \Big|_{m=0} -m (\tilde{\xi}_{\mu}\psi^{\mu}\psi^{10}+\xi^{10})\ket{0} \;.
 \end{split}
 \ee
We read off the $m$-deformed gauge transformations which 
are modified on $C_{\mu10}$, 
 \be
  \delta_{\tilde{\xi}}C_{\mu 10} \ = \ 2\tilde{\xi}^{}_{[\mu}F^{}_{10],m=0}-m\tilde{\xi}_{\mu}
  \ = \ 2\tilde{\xi}^{}_{[\mu}F^{}_{10]} \;,  
 \ee 
and on $C^{(0)}$
 \be\label{Cnottrans}
   \delta_{\xi}C^{(0)} \ = \ \xi^{j}\partial_{j}C^{(0)}-m\xi^{10} \ = \ \xi^{j}F_{j}\;,
 \ee
where we used (\ref{10thcompF}) for both equations in the last step.    
Thus, the nine-component parameter $\tilde{\xi}_{\mu}$ acts as a 
St\"uckelberg symmetry on the off-diagonal RR 2-form components, 
while the 10th diffeomorphism parameter $\xi^{10}$ acts as a  St\"uckelberg symmetry on the RR 0-form.   
The field strength of $C_{\mu 10}$ read off from (\ref{FcompIIB}), 
 \be\label{Fieldsoff}
  \widehat{F}_{\mu\nu10} \ = \ 2\partial_{[\mu}C_{\nu]10}+mb_{\mu\nu}
  +\partial_{10}C_{\mu\nu}-b_{\mu\nu}\partial_{10}C^{(0)}
  -2b_{10[\mu}\,\partial_{\nu]}C^{(0)}\;,
 \ee 
is invariant under the $\tilde{\xi}_{\mu}$ shift symmetry. Moreover, (\ref{10thcompF}) is 
invariant under $\xi^{10}$, i.e., the theory is diffeomorphism invariant 
under $x^{10}\rightarrow x^{10}-\xi^{10}(x)$ and (\ref{Cnottrans}), 
 \be
  \delta_{\xi^{10}}F^{(1)} \ = \  -md\xi^{10}+md\xi^{10} \ = \ 0 \;. 
 \ee
Thus, despite the non-covariant formulation that treats the 10th direction on a different 
footing, the theory is still fully diffeomorphism invariant, as it should be 
in view of the fact that it results from standard type IIB by the redefinition (\ref{IIBredf}). 
Since this invariance under non-covariant diffeomorphisms is somewhat unconventional, let
us also verify this for the component form given in (\ref{10thcompF}), 
 \be
 \begin{split}
  \delta_{\xi}F_{i} \ &= \ \partial_{i}\big(\xi^{j}\partial_{j}C^{(0)}-m\xi^{10}\big) \ = \ 
  \xi^{j}\partial_{j}\big(\partial_{i}C^{(0)}\big)+\partial_{i}\xi^{j}\,\partial_{j}C^{(0)}
  -m\partial_{i}\xi^{10}\\
  \ &= \ \xi^{j}\partial_{j}F_{i}+\partial_{i}\xi^{j}\big(\partial_{j}C^{(0)}-m\delta_{j}{}^{10}\big)
  \ = \  \xi^{j}\partial_{j}F_{i}+\partial_{i}\xi^{j}F_{j}\;.
 \end{split}
 \ee 
Thus, the $m$-deformed field strength transforms under the $m$-deformed 
diffeomorphisms  (\ref{Cnottrans}) with the usual Lie derivative of a 1-form 
field strength. Therefore, the action and duality relations build with this field strength 
are diffeomorphism invariant.  

To summarize, 
we have identified the 10-dimensional theory that is the T-dual to massive type IIA and 
that can be seen as a `massive' formulation of type IIB. It is unconventional in that the 10-dimensional diffeomorphism symmetry is not 
realized in the usual way, but non-linearly in the 10th direction. This is, however, 
analogous to the deformation of the gauge transformation of $C^{(1)}$ under the $b$-field gauge 
parameter in massive 
type IIA, and since the diffeomorphisms and  
$b$-field gauge symmetries are on the same footing in double field theory this result is 
not surprising.

Let us now discuss the physical content. 
We can choose a gauge for the $\tilde{\xi}_{\mu}$ St\"uckelberg symmetries by setting $C_{\mu10}=0$. 
From (\ref{Fieldsoff}) we then infer that their kinetic terms give mass terms for the 
9-dimensional components of the $b$-field, rendering these components massive. 
This is analogous to massive type IIA, but in the latter case the full 10-dimensional 
$b$-field becomes massive, carrying 36 massive degrees of freedom, while here only 
the 9-dimensional components become massive, carrying 28 massive degrees of freedom.
It turns out that the 8 missing degrees of freedom are carried instead by the Kaluza-Klein 
vector field. In order to see this, let us perform a Kaluza-Klein decomposition 
of the kinetic term involving $C^{(0)}$ (but we stress that we are not performing a 
reduction in that the fields still depend on all 10 coordinates). The standard Kaluza-Klein 
decomposition of the (inverse) metric reads 
 \be
  g^{ij} \ = \  \begin{pmatrix}
    \gamma^{\mu\nu} & - A^{\mu}   \\  - A^{\nu} & \ell^{-1}+ A^{\rho}A_{\rho} \end{pmatrix}\,,
 \ee
where $\gamma_{\mu\nu}$ denotes the 9-dimensional metric, 
$A_{\mu}$ is the Kaluza-Klein vector and $\ell$ the Kaluza-Klein scalar.  
If we choose a gauge for the $\xi^{10}$ St\"uckelberg symmetry by setting $C^{(0)}=0$, 
we infer with (\ref{10thcompF}) that the relevant term in the Lagrangian reads 
 \be\label{KKmass}
  {\cal L} \ = \ -\frac{1}{4}\sqrt{g}\,
  g^{ij}F_{i}^{}F_{j}^{} \ = \ -\frac{1}{4}\sqrt{g} g^{10,10}F_{10}^{}  F_{10}^{} \ = \ 
  -\frac{1}{4} m^2 \sqrt{\gamma}\sqrt{\ell}\left(\ell^{-1}+ A^{\mu}A_{\mu}\right)\;.
 \ee
Therefore, the Kaluza-Klein vector receives a mass term and so becomes massive by `eating'  
the RR scalar  $C^{(0)}$, thus carrying 8 massive degrees of freedom.

We have to point out that the above analysis of the physical content was somewhat naive.  
In fact, one may wonder why this theory, if obtained from massless type IIB by the mere redefinition 
(\ref{IIBredf}), exhibits a spectrum that is rather different from the usual physical content of  
type IIB, e.g., with (parts of) the $b$-field becoming massive and a cosmological term in (\ref{KKmass}). 
The point is that such a classification of the masses of various fields is 
only meaningful with respect to a particular background.  Type IIB admits a 
10-dimensional Minkowski solution, with all field strengths zero in the background, and 
it is with respect to this background that the $b$-field is massless.  Now, after the redefinition 
(\ref{IIBredf}) the theory of course still admits the same Minkowski vacuum, but now we have to 
switch on a `background flux' in order to realize this solution, 
 \be
  \vev{g_{ij}} \ = \ \eta_{ij}\;, \qquad \vev{dC^{(0)}} \ = \ mdx^{10}\;,
 \ee
because only then we have $\vev{\widehat{F}}=0$ in the Einstein equations, as follows with 
(\ref{10thcompF}). Around this background, the $b$-field is still massless.   

Thus, there is no conflict of our above analysis of `massive' type IIB 
with the usual way type IIB is presented.  The presence of massive fields 
just means that the background 
space-time we consider is \textit{not} flat space, but rather a background that is appropriate for 
the comparison to the T-dual massive type IIA.  
In fact, massive type IIA does not admit a Minkowski (or AdS) vacuum, but 
instead the D8-brane
solution that is invariant under the 9-dimensional Poincar\'e group corresponding to 
its world-volume \cite{Bergshoeff:1996ui}. 
The T-dual configuration is the D7-brane 
solution of type IIB, which is only invariant under the 8-dimensional Poincar\'e 
group \cite{Bergshoeff:1996ui}, 
and the above analysis has to be understood with respect to such a background.

Let us close this section  by comparing our result with the usual story  
that relates massive type IIA to the Scherk-Schwarz reduction 
of massless type IIB \cite{Bergshoeff:1996ui,Lavrinenko:1999xi}. 
In Scherk-Schwarz reduction one allows some fields to depend non-trivially
on the internal coordinates in such a way that this dependence drops out 
in the effective lower-dimensional theory. For the Scherk-Schwarz reduction of 
type IIB to nine dimensions relevant for T-duality, 
the Kaluza-Klein ansatz allows for a linear $x^{10}$
dependence for the RR scalar $C^{(0)}$, 
 \be\label{SSansatz}
  C^{(0)}(x^{\mu},x^{10}) \ = \ c^{(0)}(x^{\mu})-mx^{10}\;,
 \ee
where $c^{(0)}$ denotes the nine-dimensional field.  
For all other fields the ansatz is as for circle reductions, i.e., the fields 
are simply assumed to be independent of $x^{10}$. In the resulting 
action the dependence on $x^{10}$ drops out, leaving 
a massive deformation of the usual circle reduction of type IIB. 

Instead of this Scherk-Schwarz reduction one may first perform the redefinition 
(\ref{IIBredf}) and then employ a standard reduction, as is apparent 
by comparing (\ref{SSansatz}) with (\ref{IIBredf}). We conclude that the 
Scherk-Schwarz reduction of massless type IIB gives the same 9-dimensional
theory as the conventional reduction of the `massive'  formulation of type IIB.  
Thus, our results are consistent 
with \cite{Bergshoeff:1996ui,Lavrinenko:1999xi}, and the formulation of type IIB that  
appears naturally in double field theory is already adapted to the Scherk-Schwarz 
reduction.\footnote{We thank Eric Bergshoeff for 
discussions on this point.}

\section{Concluding remarks}\setcounter{equation}{0}
In this paper we have shown that the type II double field theory defined by (\ref{totaction}) and
(\ref{dualityintro}) can be extended by slightly 
relaxing the constraint (\ref{ODDconstr}) such that the RR fields may depend simultaneously 
on all 10-dimensional space-time coordinates and linearly on the 
winding coordinates. In case that only the RR 1-form carries such a dependence,  
the double field theory reduces precisely to the massive type IIA theory.
We have shown that the T-dual configuration corresponds to the case that the RR 2-form 
(\ref{2Form}) 
of type IIB carries such a dependence. This gives rise to a `massive' version 
of type IIB, whose circle reduction to nine dimensions yields the same 
theory as the Scherk-Schwarz reduction of conventional type IIB. 
This massive formulation of type IIB is still invariant under 
10-dimensional diffeomorphisms, with 
the 10th diffeomorphism  being deformed by the mass parameter.

Here we have only considered a non-trivial $\tilde{x}$ dependence for the RR 1-form
of type IIA and, in the T-dual situation, for the RR 2-form of type IIB. It is also 
consistent with the relaxed constraint to have a linear $\tilde{x}$ dependence for all higher 
RR forms $C^{(p)}$.
Such an ansatz would lead to a multiple parameter family of `massive' and non-covariant 
type II theories, but as for the type IIB case discussed above all of these deformations 
can be absorbed into a redefinition of the lower RR form $C^{(p-2)}$, as in (\ref{IIBredf}). 
The only exception is the RR 1-form which is distinguished because it leads 
to a \textit{covariant} massive deformation in that it merely deforms  the 0-form 
field strength by the scalar mass parameter. This deformation cannot be absorbed 
into a redefinition, precisely because there is no `$(-1)$-form' potential available that 
could be redefined. This is in agreement with the fact that no massive deformations
of maximal 10-dimensional supergravity are known besides massive type IIA. 
Even though $(-1)$-forms do not exist in the usual framework of  
differential geometry (such that they appear in conventional type II theories at most in a 
very formal sense),
the type II double field theory suggests a natural and concrete interpretation: $(-1)$-forms 
are 1-forms depending on the dual coordinates.

Finally, let us stress that so far we have only established the existence of the bosonic 
sector of the massive type II theories 
since the supersymmetric extension of type II double field theory has not yet been
constructed, but we strongly suspect that it exists. In fact, the approach of Siegel, 
which is equivalent to the NS-NS sector of double field theory, is already 
formulated in superspace, thereby establishing $N=1$ supersymmetry  \cite{Siegel:1993th}.
Moreover, the recent work \cite{Coimbra:2011nw} presents a rewriting of the $N=2$ 
fermionic terms and supersymmetry variations in the context of generalized 
geometry. While this does not yet prove the existence of an $N=2$ supersymmetric  
extension of double field theory, since the coordinates are not doubled
in generalized geometry, 
it provides strong evidence. These matters are currently under investigation.

\section*{Acknowledgments}
We would like to thank E.~Bergshoeff, C.~Hull, and P.~Townsend for helpful discussions and correspondence. We are especially indebted to B.~Zwiebach for numerous comments and for 
carefully reading the manuscript. 

This work is supported by the U.S. Department of Energy (DoE) under the cooperative
research agreement DE-FG02-05ER41360. The work of OH is supported by the DFG -- The German
 Science Foundation, and the work of SK is supported in part by a Samsung Scholarship.

\appendix

\section{Proof of the gauge algebra}
\setcounter{equation}{0}
Here we prove that the gauge transformations (\ref{finalgauge}) close according to (\ref{xiclosure}), 
using only the weaker constraint (\ref{weakconstr}). 
We compute 
 \be\label{gaugecomm}
 \begin{split}
  \big[ \delta_{\xi^1}, \delta_{\xi^2}\big]\chi \ &= \ \delta_{\xi^1}\big(\tfrac{1}{2}\Gamma^{N}\Gamma^{M}
  \xi_{2N}\partial_{M}\chi\big)-(1\leftrightarrow 2) \\
  \ &= \ \frac{1}{4}\Gamma^{N}\Gamma^{M}\Gamma^{P}\Gamma^{Q}\xi_{2N}\partial_{M}
  \big(\xi_{1P}\,\partial_{Q}\chi\big)-(1\leftrightarrow 2) \;.
 \end{split}
 \ee
Let us work out structures with $\partial\chi$ and $\partial^2\chi$ separately. 
Consider 
 \be\label{step001}
 \begin{split}
  &\frac{1}{8} \Gamma^{N}\Gamma^{M}\Gamma^{P}\Gamma^{Q} \xi_{2N}\xi_{1P}\partial_{M}\partial_{Q}\chi
  \ = \ \frac{1}{8}\big(\Gamma^{P}\Gamma^{Q}\Gamma^{N}\Gamma^{M}
   +[\Gamma^{N}\Gamma^{M},\Gamma^{P}\Gamma^{Q}]\big)\xi_{2N}\xi_{1P}\partial_{M}\partial_{Q}\chi\\
   \ &= \  \frac{1}{8}\left(\Gamma^{P}\Gamma^{Q}\Gamma^{N}\Gamma^{M}
   -2\big(\eta^{MP}\Gamma^{QN}+\eta^{NQ}\Gamma^{PM}\big)\right) 
    \xi_{2N}\xi_{1P}\partial_{M}\partial_{Q}\chi\;,
 \end{split}
 \ee
where we used the constraint (\ref{weakconstr}) for $\chi$ and the symmetry in $M,Q$. 
If we antisymmetrize 
in $1\leftrightarrow 2$ 
and use the symmetry in $M,Q$, we infer that the term on the left-hand side in the first line 
is minus the first term in the second line. 
Similarly, upon antisymmetrization $1\leftrightarrow 2$, the final two terms in the second line 
are equal. Rearranging terms, we thus get 
 \be\label{step002}
 \begin{split}
  \frac{1}{4} &\Gamma^{N}\Gamma^{M}\Gamma^{P}\Gamma^{Q} \xi_{2N}\xi_{1P}\partial_{M}\partial_{Q}\chi
  -(1\leftrightarrow 2) \ = \ 
  \frac{1}{2}\Gamma^{M}\Gamma^{P}\xi_2^{N}\xi_{1P}\partial_M\partial_N\chi-(1\leftrightarrow 2)\\
 \ &= \ \frac{1}{2}\Gamma^{M}\partial_{M}\big(\xi_2^{N}\Gamma^{P}\xi_{1P}\partial_{N}\chi\big)
 -\frac{1}{2}\Gamma^{M}\Gamma^{P}\partial_{M}\big(\xi_2^{N}\xi_{1P}\big)\partial_{N}\chi-(1\leftrightarrow 2)\;,
 \end{split} 
 \ee
where we used in the first equality that a term with $\eta^{MP}$ is zero by the 
antisymmetry   $1\leftrightarrow 2$. The first term in the second line takes the form of a field-dependent
$\lambda$ gauge transformation. 

We turn now to the terms proportional to $\partial\chi$ in (\ref{gaugecomm}),  
 \be\label{step003}
  \frac{1}{4}\Gamma^{N}\Gamma^{M}\Gamma^{P}\Gamma^{Q}\xi_{2N}\partial_{M}\xi_{1P}
  \partial_{Q}\chi \ = \ 
  \frac{1}{2}\Gamma^{P}\Gamma^{Q}\xi_{2}^{M}\partial_{M}\xi_{1P}\partial_{Q}\chi
  -\frac{1}{4}\Gamma^{M}\Gamma^{N}\Gamma^{P}\Gamma^{Q}\xi_{2N}\partial_{M}\xi_{1P}\partial_{Q}\chi\;,
 \ee
where we used the Clifford algebra for $\Gamma^{N}\Gamma^{M}$. 
The first term on the right-hand side 
takes the form of a $\xi^{M}$ gauge transformation (\ref{finalgauge}). The second term can be 
re-written as 
 \be\label{step004}
 \begin{split}
   -\frac{1}{4}&\Gamma^{M}\Gamma^{N}\Gamma^{P}\Gamma^{Q}\xi_{2N}\partial_{M}\xi_{1P}\partial_{Q}\chi
   \ = \  -\frac{1}{4}\Gamma^{M}\big(\Gamma^{Q}\Gamma^{N}\Gamma^{P}+[\Gamma^{N}\Gamma^{P},
   \Gamma^{Q}]\big)\xi_{2N}\partial_{M}\xi_{1P}\partial_{Q}\chi \\
   \ &= \ -\frac{1}{4}\Gamma^{M}\Gamma^{Q}\Gamma^{N}\Gamma^{P}
   \xi_{2N}\partial_{M}\xi_{1P}\partial_{Q}\chi+\frac{1}{2}\Gamma^{M}
   \big(\eta^{QN}\Gamma^{P}-\eta^{QP}\Gamma^{N}\big) \xi_{2N}\partial_{M}\xi_{1P}\partial_{Q}\chi\;.
\end{split}
\ee   
Antisymmetrizing $1\leftrightarrow 2$,  
the 
last term in the second line gives 
 \be
 \begin{split}
  \frac{1}{2}\Gamma^{M}\Gamma^{P}\xi_{2}^{N}\partial_{M}\xi_{1P}\partial_{N}\chi
  &-\frac{1}{2}\Gamma^{M}\Gamma^{N}\xi_{2N}\partial_{M}\xi_{1}^{P}\partial_{P}\chi -(1\leftrightarrow 2)\\
   \ &= \ \frac{1}{2}\Gamma^{M}\Gamma^{P}\partial_{M}\big(\xi_2^{N}\xi_{1P}\big)\partial_{N}\chi
   -(1\leftrightarrow 2)\;, 
 \end{split}
 \ee
which cancels against the same structure in (\ref{step002}). The first term in the second line of (\ref{step004}), 
antisymmetrized in $1\leftrightarrow 2$, can be simplified as follows
 \be\label{step987}
 \begin{split}
  -\frac{1}{4}&\Gamma^{M}\Gamma^{Q}\big(\eta^{NP}+\Gamma^{NP}\big)\big(\xi_{2N}\partial_{M}\xi_{1P}
  -\xi_{1N}\partial_{M}\xi_{2P}\big)\partial_{Q}\chi \\
  \ = \  &-\frac{1}{4}\Gamma^{M}\Gamma^{Q}\Gamma^{NP}
  \partial_{M}\big(\xi_{2N}\xi_{1P}\big)\partial_{Q}\chi  
  -\frac{1}{4}\Gamma^{M}\Gamma^{Q}\big(\xi_2^{N}\partial_{M}\xi_{1N}-\xi_1^{N}\partial_{M}\xi_{2 N}\big)
 \partial_{Q}\chi\;.
 \end{split}
 \ee
The second term is of the form of a $\xi^{M}$ gauge transformation. 
Commuting gamma matrices and using the weak constraint, the first term can be 
rewritten as a $\lambda$ gauge transformation, 
 \be\label{STEP009}
 \begin{split}
  -\frac{1}{4}\Gamma^{M}\partial_{M}\Big[\big(& \Gamma^{NP}\Gamma^{Q}+2\big(\eta^{QN}\Gamma^{P}
  -\eta^{QP}\Gamma^{N}\big)\big)\xi_{2N}\xi_{1P}\partial_{Q}\chi\Big] \\
  \ &= \ 
   -\frac{1}{2}\slashed{\partial}\Big[\big(\slashed{\xi}_2\slashed{\xi}_1-\slashed{\xi}_1\slashed{\xi}_2\big)
  \slashed{\partial}\chi\Big] - 
  \frac{1}{2}\Gamma^{M}\partial_{M}\Big[\big(\xi_2^{N}\Gamma^{P}\xi_{1P}
  -\xi_1^{N}\Gamma^{P}\xi_{2P}\big)\partial_{N}\chi\Big]\;.
 \end{split}
 \ee
The second term in here cancels against the the first term in (\ref{step002}). 

Summarizing, the surviving structures are the $\xi^{M}$ gauge transformations in 
(\ref{step003}) and (\ref{step987}) and the $\lambda$ transformation in (\ref{STEP009}), 
combining into
 \be
  \big[ \delta_{\xi^1}, \delta_{\xi^2}\big]\chi \ = \ \frac{1}{2}\Gamma^{P}\Gamma^{Q}\Big( 
  \xi_2^{M}\partial_{M}\xi_{1P}-\frac{1}{2}\xi_2^{M}\partial_{P}\xi_{1M}-(1\leftrightarrow 2)\Big)
  \partial_{Q}\chi
  +\slashed{\partial}\lambda\;,
 \ee
with the $\lambda$ parameter 
(\ref{lambdaparam}). Thus, the gauge algebra closes as stated in (\ref{xiclosure}).

\end{document}